\documentclass[twocolumn,showpacs,aps,preprintnumbers,amsmath,amssymb,superscriptaddress,footinbib]{revtex4}

\usepackage{graphicx,subfigure,multirow}
 \usepackage{longtable}
\usepackage{dcolumn}
\usepackage{bm}
\usepackage[all]{xy}
\usepackage{color}
\usepackage[usenames,dvipsnames]{xcolor}
\usepackage[unicode=true,pdfusetitle,
 bookmarks=true,bookmarksnumbered=false,bookmarksopen=false,citecolor=Turquoise,
 breaklinks=false,pdfborder={0 0 1},backref=false,colorlinks=true,pdfpagemode=FullScreen]
 {hyperref}

\usepackage{ulem}

\makeatletter

\newcommand{\fmslash}[2][0mu]{%
  \mathchoice
    {\fmsl@sh\displaystyle{#1}{#2}}%
    {\fmsl@sh\textstyle{#1}{#2}}%
    {\fmsl@sh\scriptstyle{#1}{#2}}%
    {\fmsl@sh\scriptscriptstyle{#1}{#2}}}
\newcommand{\fmsl@sh}[3]{%
  \m@th\ooalign{$\hfil#1\mkern#2/\hfil$\crcr$#1#3$}}
\makeatother

\newcommand{\lsim}{{\;\raise0.3ex\hbox{$<$\kern-0.75em\raise-1.1ex\hbox{$\sim$}}\;}}
\newcommand{\gsim}{{\;\raise0.3ex\hbox{$>$\kern-0.75em\raise-1.1ex\hbox{$\sim$}}\;}}

\newcommand{\beq}{\begin{equation}}
\newcommand{\eeq}{\end{equation}}
\newcommand{\bea}{\begin{eqnarray}}
\newcommand{\eea}{\end{eqnarray}}
\mathchardef\minus="002D

\newcommand{\met}{{\fmslash E_T}}
\usepackage{color}

\begin{document}
\title{The 750 GeV Diphoton Excess May Not Imply a 750 GeV Resonance}

\author{Won Sang Cho}
\affiliation{Center for Theoretical Physics of the Universe, Institute for Basic Science (IBS), Daejeon, 34051, Korea}
\author{Doojin Kim}
\affiliation{Department of Physics, University of Florida, Gainesville, FL 32611, USA}
\author{Kyoungchul Kong}
\affiliation{Department of Physics and Astronomy, University of Kansas, Lawrence, KS 66045, USA}
\author{Sung Hak Lim}
\affiliation{Center for Theoretical Physics of the Universe, Institute for Basic Science (IBS), Daejeon, 34051, Korea}
\affiliation{Department of Physics, KAIST, 291 Daehak-ro, Yuseong-gu, Daejeon, 34141, Korea}
\author{Konstantin T. Matchev}
\affiliation{Department of Physics, University of Florida, Gainesville, FL 32611, USA}
\author{Jong-Chul Park}
\affiliation{Department of Physics, Chungnam National University, Daejeon 305-764, Korea}
\author{Myeonghun Park}
\affiliation{Center for Theoretical Physics of the Universe, Institute for Basic Science (IBS), Daejeon, 34051, Korea}
\date{\today}

\preprint{
\begin{minipage}[b]{1\linewidth}
\begin{flushright}
CTPU-15-27\\
 \end{flushright}
\end{minipage}
}

\begin{abstract}
We discuss non-standard interpretations of the $750$ GeV diphoton excess recently reported by the ATLAS and CMS Collaborations
which do {\it not} involve a new, relatively broad, resonance with a mass near $750$ GeV.
Instead, we consider the sequential cascade decay of a much heavier, possibly quite narrow, 
resonance into two photons along with one or more additional particles. 
The resulting diphoton invariant mass signal is generically rather broad, as suggested by the data.
We examine three specific event topologies --- the ``antler'', the ``sandwich'', and the 2-step cascade decay,
and show that they all can provide a good fit to the observed published data. In each case, we
delineate the preferred mass parameter space selected by the best fit.
In spite of the presence of extra particles in the final state, the measured diphoton $p_T$ spectrum
is moderate, due to its anti-correlation with the diphoton invariant mass.
We comment on the future prospects of discriminating with higher statistics between our scenarios,
as well as from more conventional interpretations.
\end{abstract}


\maketitle

\paragraph*{{\bf Introduction.}}
Recently, the ATLAS and CMS Collaborations have reported first results with data obtained
at the Large Hadron Collider (LHC) operating at 13 TeV. The data shows an intriguing
excess in the inclusive diphoton final state \cite{ATLAS, CMS:2015dxe}.
The ATLAS Collaboration further reported that about 15 events in the diphoton invariant mass distribution are observed above the
Standard Model (SM) expectation at 3.9$\sigma$ local significance (2.3$\sigma$ global significance) with 3.2 fb$^{-1}$ of data.
The excess appears as a bump at $M\sim$ 750 GeV with a relatively broad width $\Gamma\sim$ 45 GeV,
resulting in $\Gamma/M \sim 0.06$ \cite{ATLAS}. Similar results are reported by the CMS Collaboration for 2.6 fb$^{-1}$ of data ---
there are about 10 excess events at a local significance of 2.6$\sigma$ (2.0$\sigma$) assuming a narrow (wide) width \cite{CMS:2015dxe}.
The anomalous events are not accompanied by significant extra activity, e.g.~missing transverse energy $\met$ \cite{Moriond}.
The required cross section for the excess is $\sim $ 10 fb at 13 TeV, and so far no indication of a similar
excess has been observed in other channels.

While waiting for the definitive verdict on this anomaly from additional LHC data, it is fun to speculate on
new physics scenarios which are consistent with the current data.
Since the excess was seen in the diphoton invariant mass spectrum, the most straightforward
interpretation would involve the production of a resonance with mass near 750 GeV, which decays directly
to two photons. The relative broadness of the observed feature would imply that this resonance
has a relatively large width, creating some tension with its non-observation in other channels.
Since the initial announcement, many models along those lines have been proposed, e.g.~in
the context of extended Higgs sectors \cite{higgs}, supersymmetry \cite{susy}, 
extra dimensions \cite{ed},  strong dynamics \cite{sd},
or effective field theory \cite{eft}.

\begin{figure}[t]
\centering
\includegraphics[width=8cm]{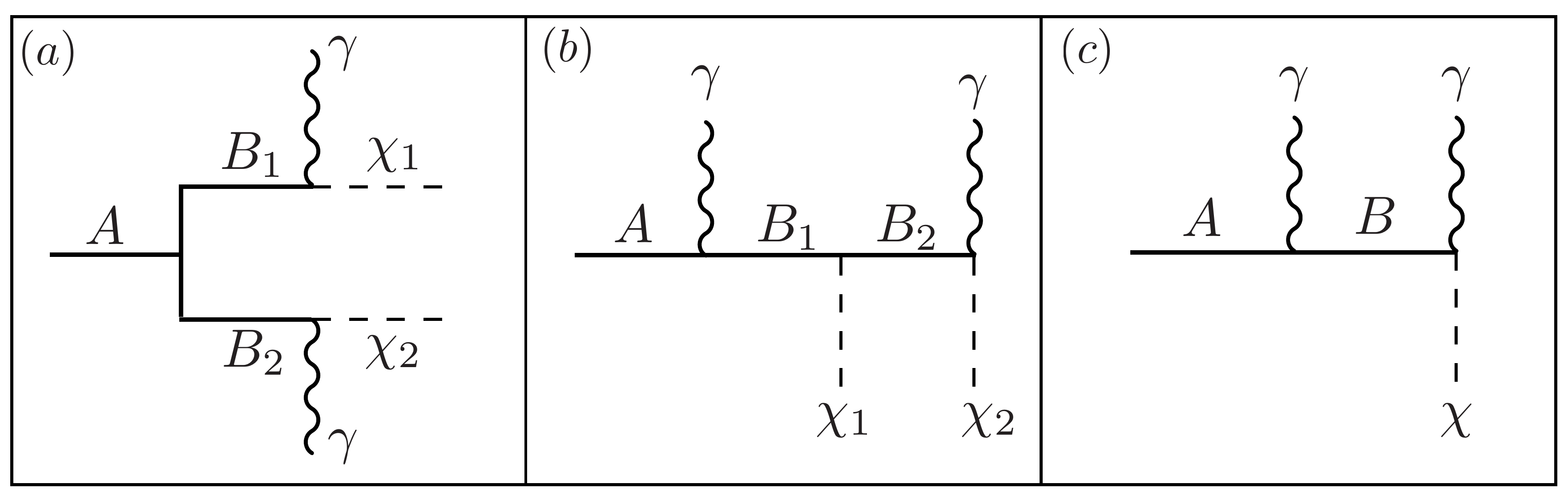}
\caption{\label{fig:diagram}
The event topologies with two photons $\gamma$ (wavy lines) and up to two additional particles $\chi_i$ (dashed lines) 
under consideration in this letter: (a) antler,  (b) sandwich, and (c) 2-step cascade decay.
Solid lines correspond to heavier resonances ($A$,  $B_i$). }
\end{figure}

In this letter, we entertain a different interpretation of the diphoton excess in the context of a
sequential cascade decay of a much heavier, possibly quite narrow, resonance, resulting in a final state with two photons and
one or two {\it additional} particles (see also \cite{Knapen:2015dap}). Three specific examples of such simplified 
model \footnote{Explicit models resulting in such signatures can be readily constructed --- for example, the 
antler topology of Fig.~\ref{fig:diagram}$(a)$ may result from the production of a level 2 Kaluza-Klein (KK) partner
decaying to two level 1 KK particles \cite{Datta:2005zs} in a Universal Extra Dimensions scenario with 
KK number violation \cite{Macesanu:2002ew}. Another possibility is the decay of a Higgs boson to two 
neutralinos in models with gauge-mediated supersymmetry \cite{DiazCruz:2003bx}.}
event topologies are exhibited in Fig.~\ref{fig:diagram}:
an ``antler'' topology~\cite{Han:2009ss} in Fig.~\ref{fig:diagram}$(a)$,
a ``sandwich''  topology~\cite{Agashe:2010gt} in Fig.~\ref{fig:diagram}$(b)$
and a 2-step cascade decay in Fig.~\ref{fig:diagram}$(c)$. In such scenarios,
the resulting diphoton invariant mass $m_{\gamma\gamma}$
is typically characterized by a somewhat broad
distribution, which eliminates the necessity of an intrinsically broad resonance.
Furthermore, the peak of the $m_{\gamma\gamma}$ distribution is found near the upper kinematic endpoint,
making it likely that the first signal events will be seen at large invariant mass,
while the low mass tail remains buried under the steeply falling SM background.
Interestingly, for signal events with the required extreme values of $m_{\gamma\gamma}$,
the transverse momentum of the diphoton system $p_T^{\gamma\gamma}$ turns out to be rather moderate,
due to its anti-correlation with the diphoton mass $m_{\gamma\gamma}$.
Given the small signal statistics ($\mathcal{O}(10)$ events) such cascade decays may easily
fake the standard diphoton resonance signature, and deserve further scrutiny.
We note that this observation is not restricted to the diphoton channel, but is quite 
general and applicable to any inclusive resonance search in a two-body final state.

\paragraph*{{\bf Diphoton invariant mass spectrum.}}
We first review the diphoton invariant mass distributions
corresponding to the above-mentioned three event topologies from Fig.~\ref{fig:diagram}.
The differential distribution of the diphoton invariant mass $m\equiv m_{\gamma\gamma}$
\bea
\frac{dN}{dm}\equiv f(m;\,M_A,\, M_{B_i},\, M_{\chi_i}) \label{eq:formula}
\eea
is known analytically (see, e.g., \cite{Cho:2012er}) and is simply a function of the unknown masses
$M_A$, $M_{B_i}$ and $M_{\chi_i}$.
The kinematic endpoint (henceforth denoted as $E$) is defined as the maximum value of $m$ allowing a non-zero $f(m)$, i.e., $E\equiv \max\lbrace m \rbrace$.

Ignoring for the moment spin correlations and assuming pure phase space distributions,
the shape in the case of the antler topology of Fig.~\ref{fig:diagram}$(a)$ is given by~\cite{Han:2009ss}
\bea
f(m)\sim \left\{
\begin{array}{l l}
\eta\, m \, , & 0 \leq m \leq e^{-\eta}E, \\ [1mm]
m \ln(E/m)  \, , & e^{-\eta}E \leq m \leq E,
\end{array}\right.  \label{eq:antlerf}
\eea
where the endpoint $E$ and the parameter $\eta$ are defined in terms of the mass parameters as
\bea
E &=&  \sqrt{  e^{\eta}   (M_{B_1}^2-M_{\chi_1}^2)(M_{B_2}^2-M_{\chi_2}^2)/(M_{B_1}M_{B_2}) } \,,~~ \\
\eta &=&  \cosh^{-1} \left[ (M_A^2-M_{B_1}^2-M_{B_2}^2)/(2M_{B_1}M_{B_2}) \right].
\eea
The corresponding shape for the sandwich topology is given by the same expression
(\ref{eq:antlerf}), only this time $E$ and $\eta$ are defined as follows~\cite{Cho:2012er}:
\bea
E &=& \sqrt{  e^{\eta} (M_{A}^2-M_{B_1}^2)(M_{B_2}^2-M_{\chi_2}^2)/(M_{B_1}M_{B_2}) }\,, \\
\eta &=& \cosh^{-1} \left[ (M_{B_1}^2+M_{B_2}^2-M_{\chi_1}^2)/(2M_{B_1}M_{B_2})\right].
\eea
In both cases, 
for small enough values of $\eta$, the peak location $e^{-\eta}E$ can be arbitrarily close to the endpoint $E$.

Finally, the two-step cascade decay has the well-known triangular shape
\bea
f(m)\sim m, \label{eq:cascadef}
\eea
where the distribution extends up to
\bea
E = \sqrt{(M_A^2-M_B^2)(M_B^2-M_{\chi}^2)/M_B^2}\,.
\label{cendpoint}
\eea

For all three event topologies in Fig.~\ref{fig:diagram} 
(assuming small enough values of $\eta$ in (\ref{eq:antlerf})), the distributions are characterized by a
relatively broad peak near the kinematic endpoint, and a continuously falling tail to lower values of $m$.
Given that the SM background distribution for $m_{\gamma\gamma}$ is a
very steeply falling function, the low $m$ tail can be easily hidden in the background, and the only
feature of the signal distribution which would be visible in the early data is the peak itself.

\paragraph*{{\bf Signal models.}}
For the numerical studies below we choose the following signal models realizing the topologies of Fig.~\ref{fig:diagram}.
In the antler topology of Fig.~\ref{fig:diagram}(a), the particle $A$ ($B_i$, $\chi_i$) is a scalar (fermion, fermion),
and the fermion coupling to the photon is vector-like, $\sim \bar{B}_i \sigma^{\mu\nu}  \chi_i F_{\mu\nu}$,
where  $F_{\mu\nu}$ is the photon field strength tensor.
For the sandwich topology of Fig.~\ref{fig:diagram}(b), particle $A$ is a heavy $U(1)$ vector boson with 
field strength tensor $F'_{\mu\nu}$, which couples to a scalar $B_1$ as $\sim B_1 F'_{\mu\nu} F^{\mu\nu}$, 
while $B_2$ and $\chi_i$ are fermions with vector-like couplings to photons as before.
Finally, in the two-step cascade decay of Fig.~\ref{fig:diagram}(c), 
$A$ and $\chi$ are vector particles coupling to a scalar $B$ as above.
In all three cases, the diphoton invariant mass distribution is given by the 
analytical results of the previous section \cite{Edelhauser:2012xb}.

\begin{figure}[t!]
\centering
\includegraphics[width=8cm]{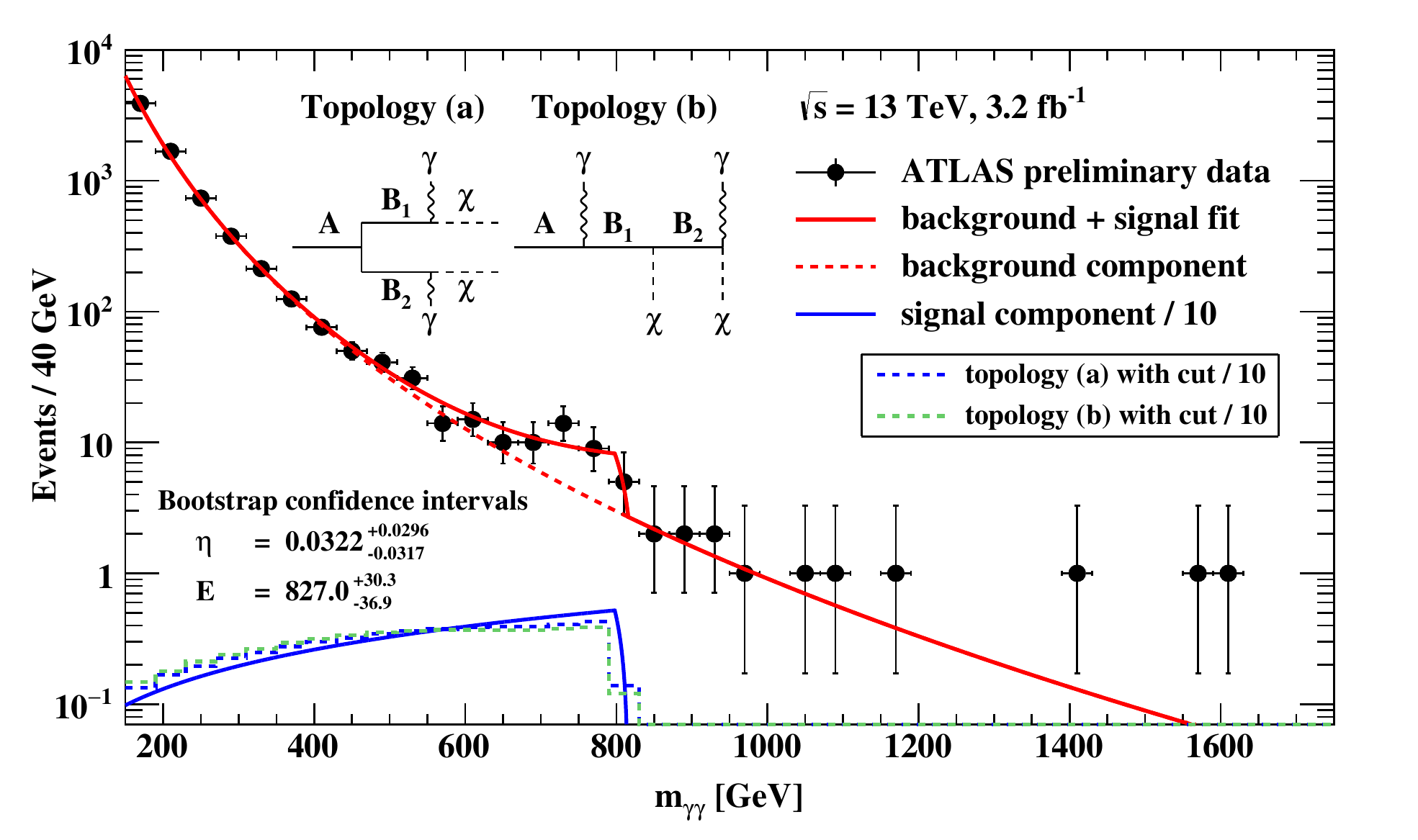} \\ \vspace{-0.14cm}
\includegraphics[width=8cm]{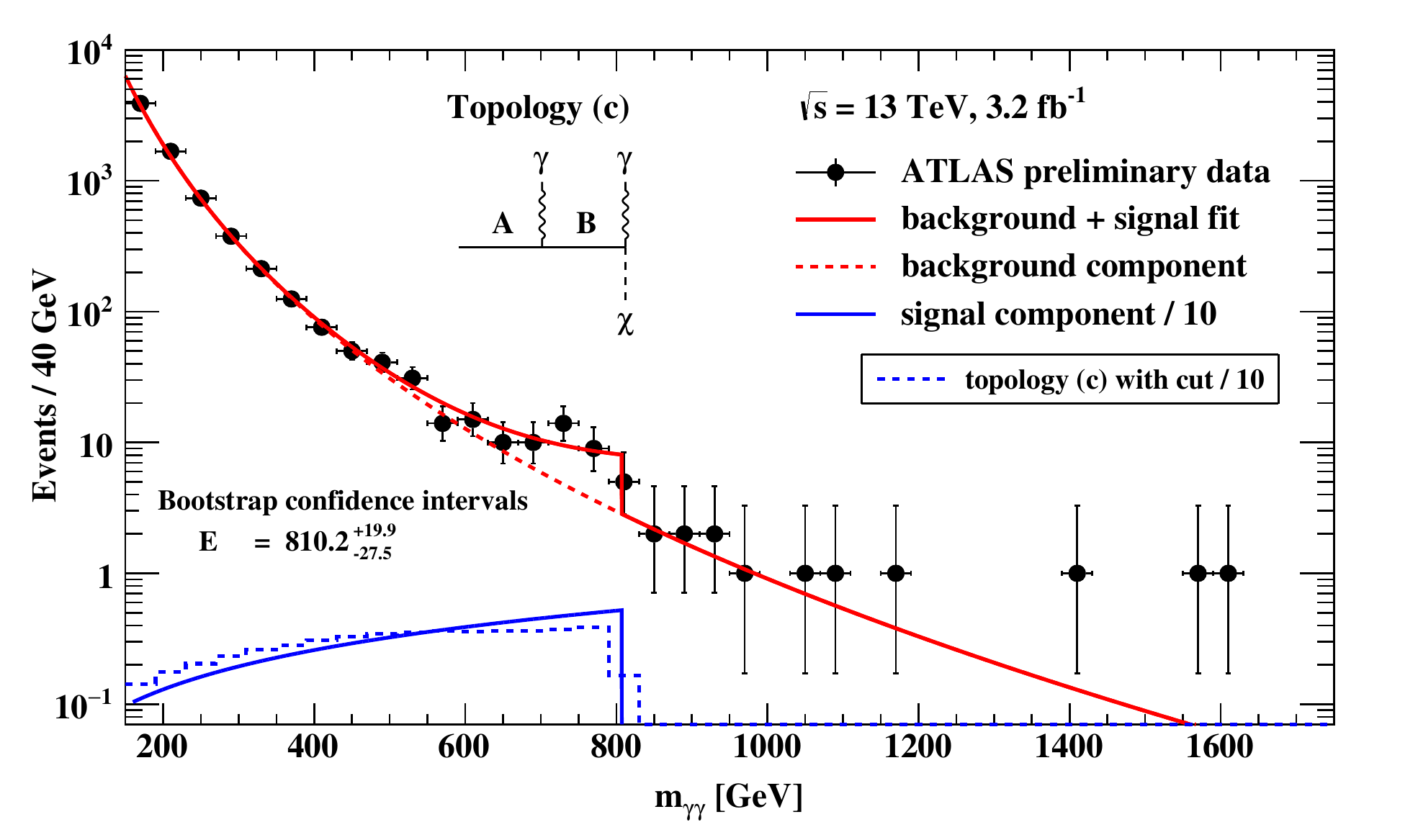}
\caption{\label{fig:fitting} Upper panel: the ATLAS diphoton data (black dots) and our fit results with the antler and
sandwich event topologies, Eq.~(\ref{eq:antlerf}).
The red solid curve represents the best-fit signal plus background distribution. 
The blue dashed (green dashed) curve represents the best-fit Monte Carlo event distribution in the antler (sandwich) case
after incorporating the ATLAS analysis cuts.
Lower panel: the same, but for the 2-step cascade decay of Fig.~\ref{fig:diagram}(c). }
\end{figure}

\paragraph*{{\bf Data analysis.}}
Given the analytical results (\ref{eq:antlerf}-\ref{cendpoint})
from the previous section, we now try to fit the three models from Fig.~\ref{fig:diagram} to
the $m_{\gamma\gamma}$ spectrum data reported by the ATLAS Collaboration~\cite{ATLAS}
(black dots in Fig.~\ref{fig:fitting}). To describe the background portion in the data, we introduce the same background model function as in~\cite{ATLAS}:
\bea
f_{bg}(x; b, a) = (1 - x^{1/3})^b x^a,
\label{eq:bg}
\eea
where $a$ and $b$ are fit parameters to be determined by data and $x = m / \sqrt{s}$ with $\sqrt{s}=13\,\text{TeV}$. We then perform likelihood fits using combined signal $+$ background templates using
$f(m)$ from Eqs.~(\ref{eq:antlerf}, \ref{eq:cascadef}) and $f_{bg}(m)$ from Eq.~\eqref{eq:bg}.
Our fit results for the case of Fig.~\ref{fig:diagram}(a-b) (Fig.~\ref{fig:diagram}(c)) are shown in the upper (lower) panel of Fig.~\ref{fig:fitting}.
The red solid curves represent the best-fit models (i.e., signal$+$background), while the red dashed (blue solid) curves describe their background (signal) components.
To estimate parameter errors more carefully with low statistics, we generate 10,000 pseudo data sets via random samplings of the data point in each bin, assuming a Poisson distribution with its mean value parameter set to the number of events in each bin reported by the ATLAS collaboration (any zero bins in the real data were always sampled to be 0).
We then conduct the same fit procedure explained above for all pseudo data sets and obtain distributions of fitted model parameters, from which we extract mean values and 1$\sigma$ confidence intervals, along with reduced $\chi^2$ distributions.

For the antler and sandwich cases, the relevant reduced $\chi^2$ distribution yields a mean (median) value of 0.98 (0.93) with the Gaussian width around 1, indicating that our fitting template accommodates pseudo data samples well enough.
The extracted best-fit parameter values and their 1$\sigma$ errors are
\bea
\eta=0.0322^{+0.0296}_{-0.0317},\quad E=827.0^{+30.3}_{-36.9}\,\,\text{GeV}.
\label{etaE_fit}
\eea
Due to the set of cuts applied in the ATLAS analysis to suppress the SM backgrounds,
the resulting signal distributions could be distorted.
In order to account for those effects,
we simulate single production of particle $A$ at LHC13
with {\verb^MadGraph5_aMC@NLO^} \cite{Alwall:2014hca}, 
followed by {\tt Pythia 6.4} \cite{Sjostrand:2006za}
and {\tt Delphes 3} \cite{deFavereau:2013fsa}.
We take  $M_A=1.7$ TeV and the remaining masses are chosen in accordance with the best-fit $E$ and $\eta$ from Eq.~(\ref{etaE_fit}). 
Since the antler and the sandwich scenarios have, in principle, different cut-sensitivity,
we show the corresponding distributions with the blue (antler) and green (sandwich) dashed curves in the upper panel of Fig.~\ref{fig:fitting}.

For the two-step cascade scenario,
the relevant reduced $\chi^2$ distribution shows a mean (median) value of 0.69 (0.67) with the Gaussian width around 0.5, indicating that this model also reproduces the data well enough.
The best-fit value for $E$ and its 1$\sigma$ error are reported as
\bea
E= 810.2^{+19.9}_{-27.5} \, \hbox{ GeV}\,.
\label{E}
\eea
As before, the signal distribution after cuts is shown by the blue dashed curve in the lower panel of Fig.~\ref{fig:fitting}.

\begin{figure}[t!]
\centering
\begin{tabular}{cc}
\includegraphics[clip=true, trim=0.1cm 0.3cm 0cm 0.cm, height=4.2cm, width=4.2cm]{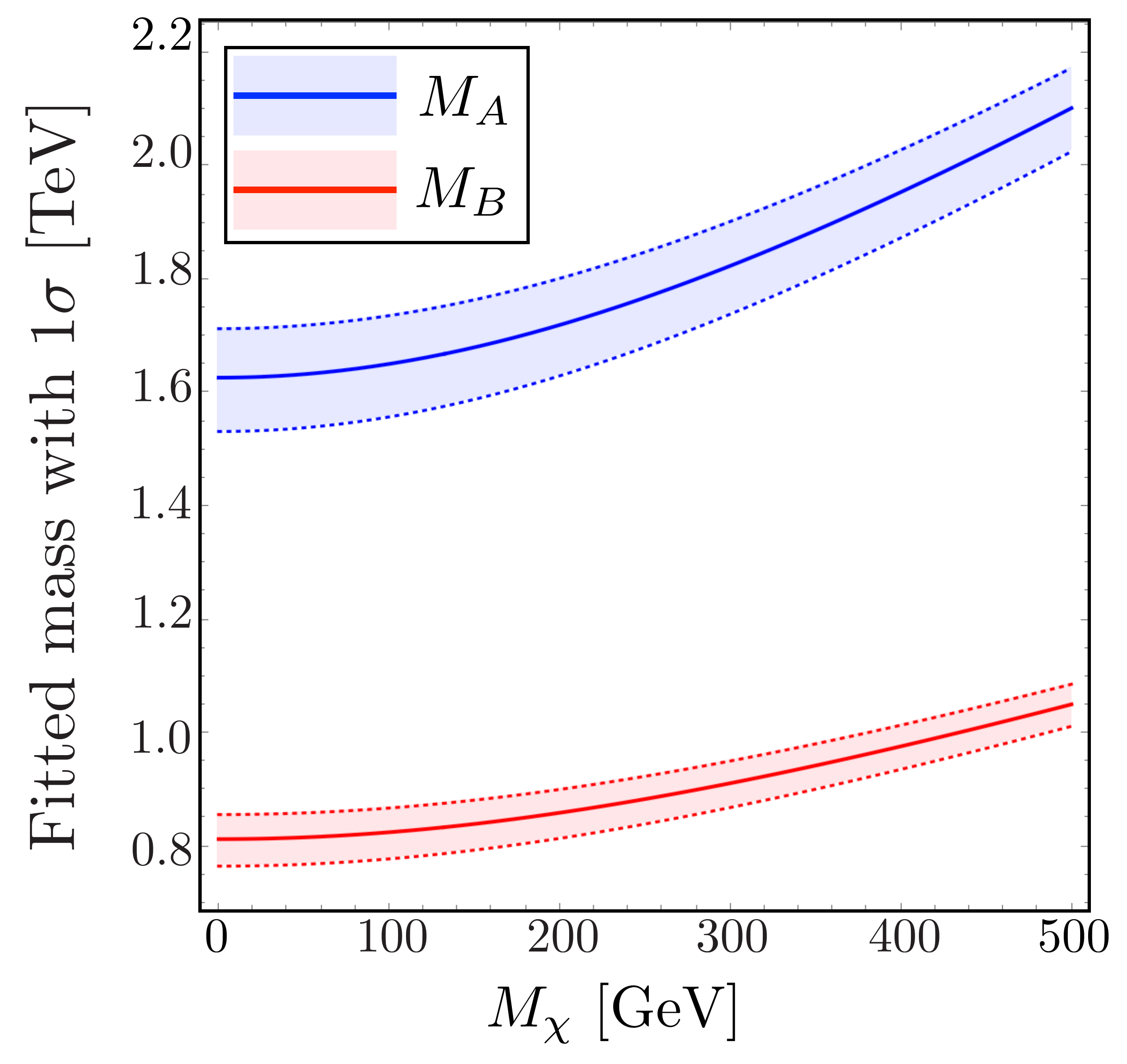} &
\includegraphics[clip=true, trim=0cm -0.1cm 0cm 0.7cm, height=4.2cm, width=4.2cm]{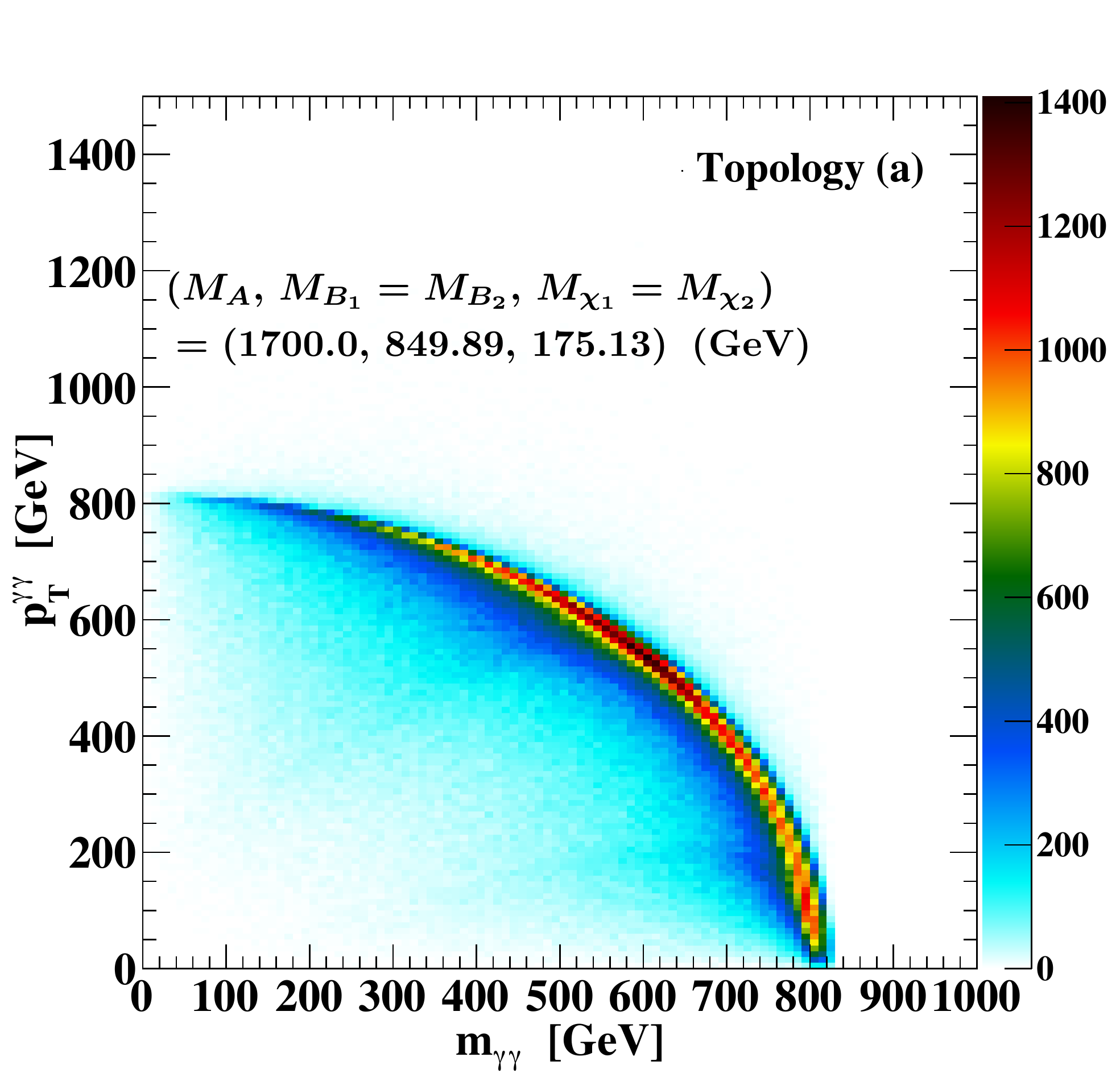} \\
\includegraphics[clip=true, trim=0.1cm 0.3cm 0cm 0.cm, height=4.2cm, width=4.2cm]{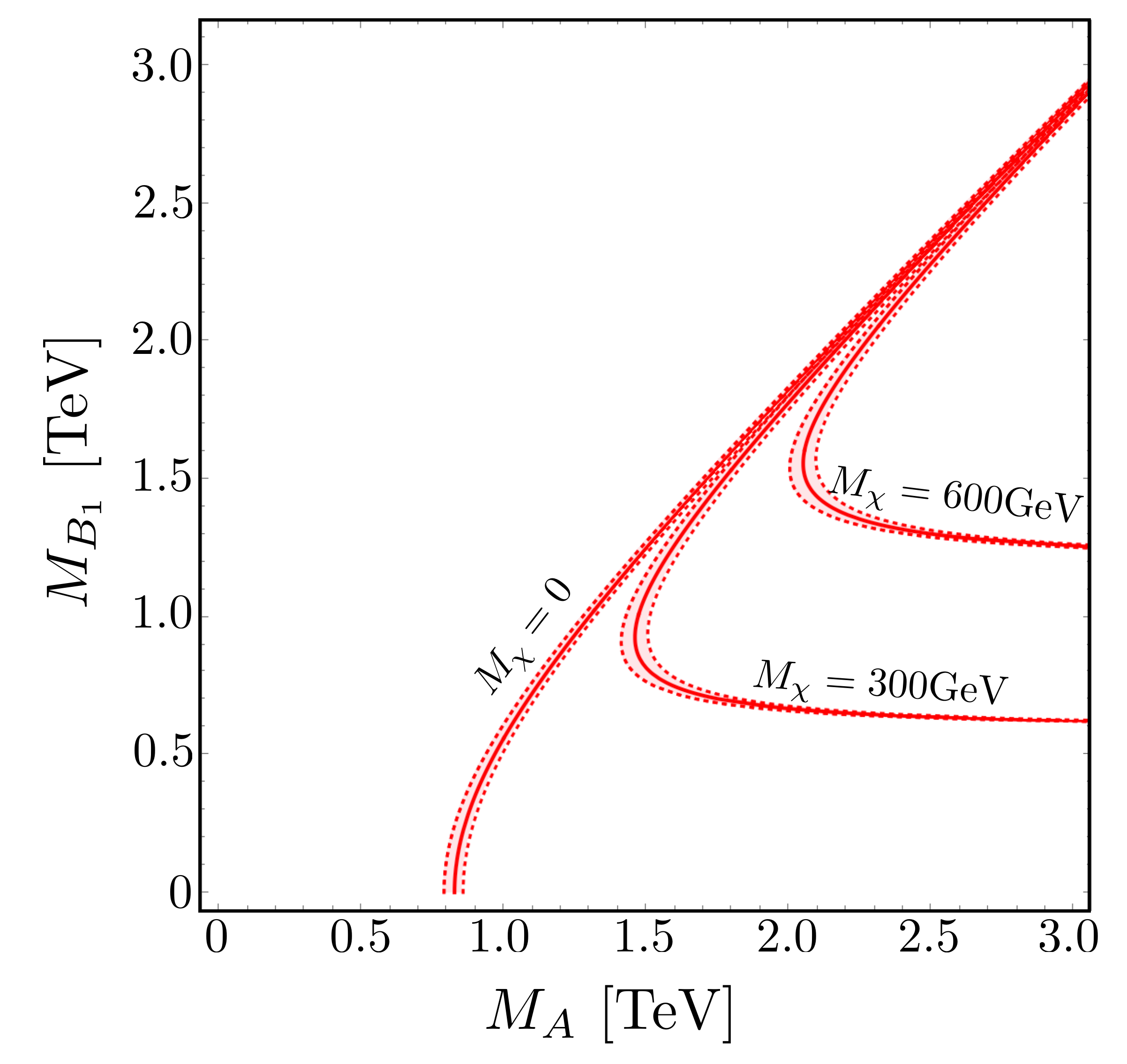} &
\includegraphics[clip=true, trim=0cm -0.1cm 0cm 0.7cm, height=4.2cm, width=4.2cm]{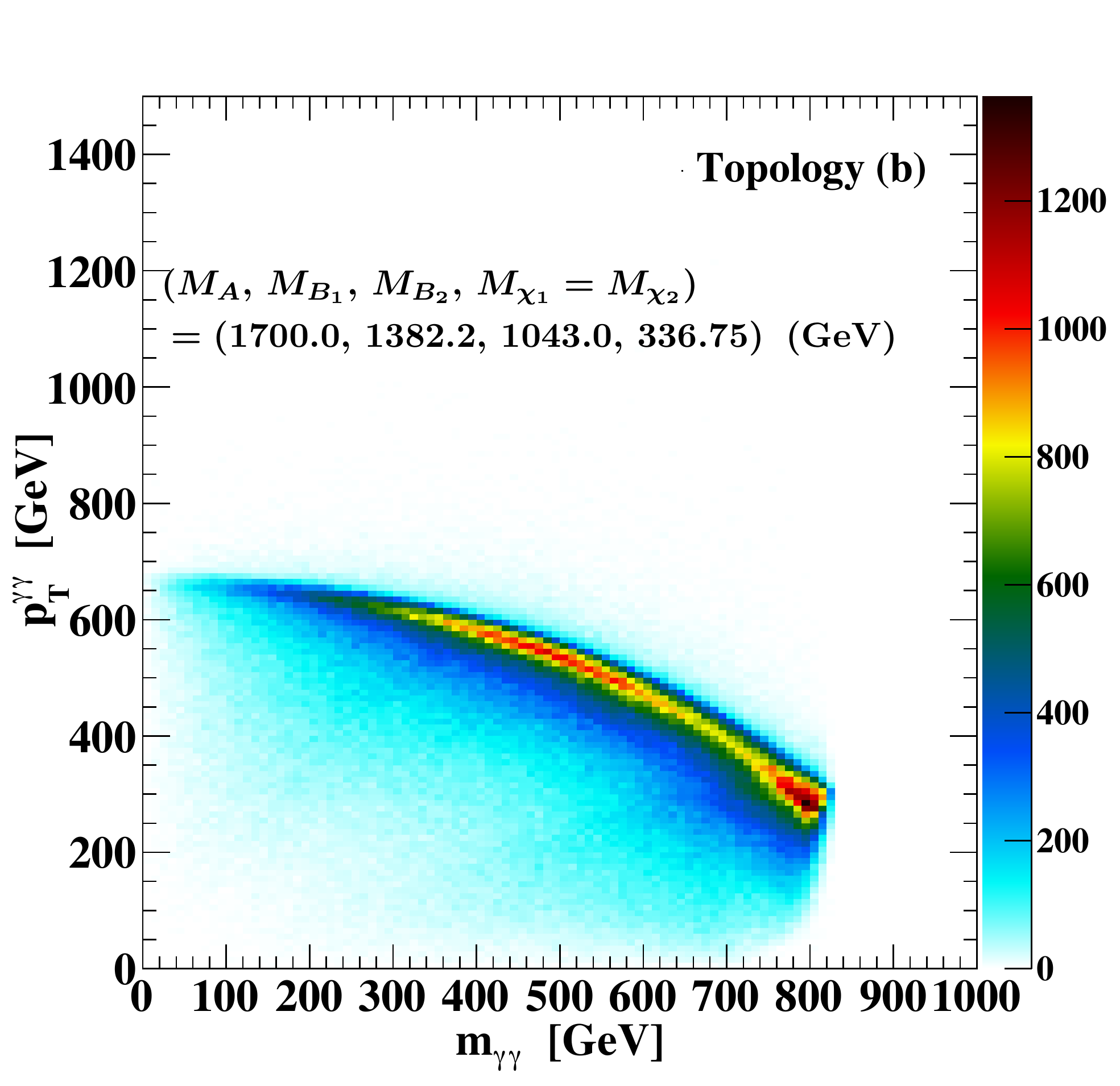} \\
\includegraphics[clip=true, trim=0.1cm 0.3cm 0cm 0cm, height=4.2cm, width=4.2cm]{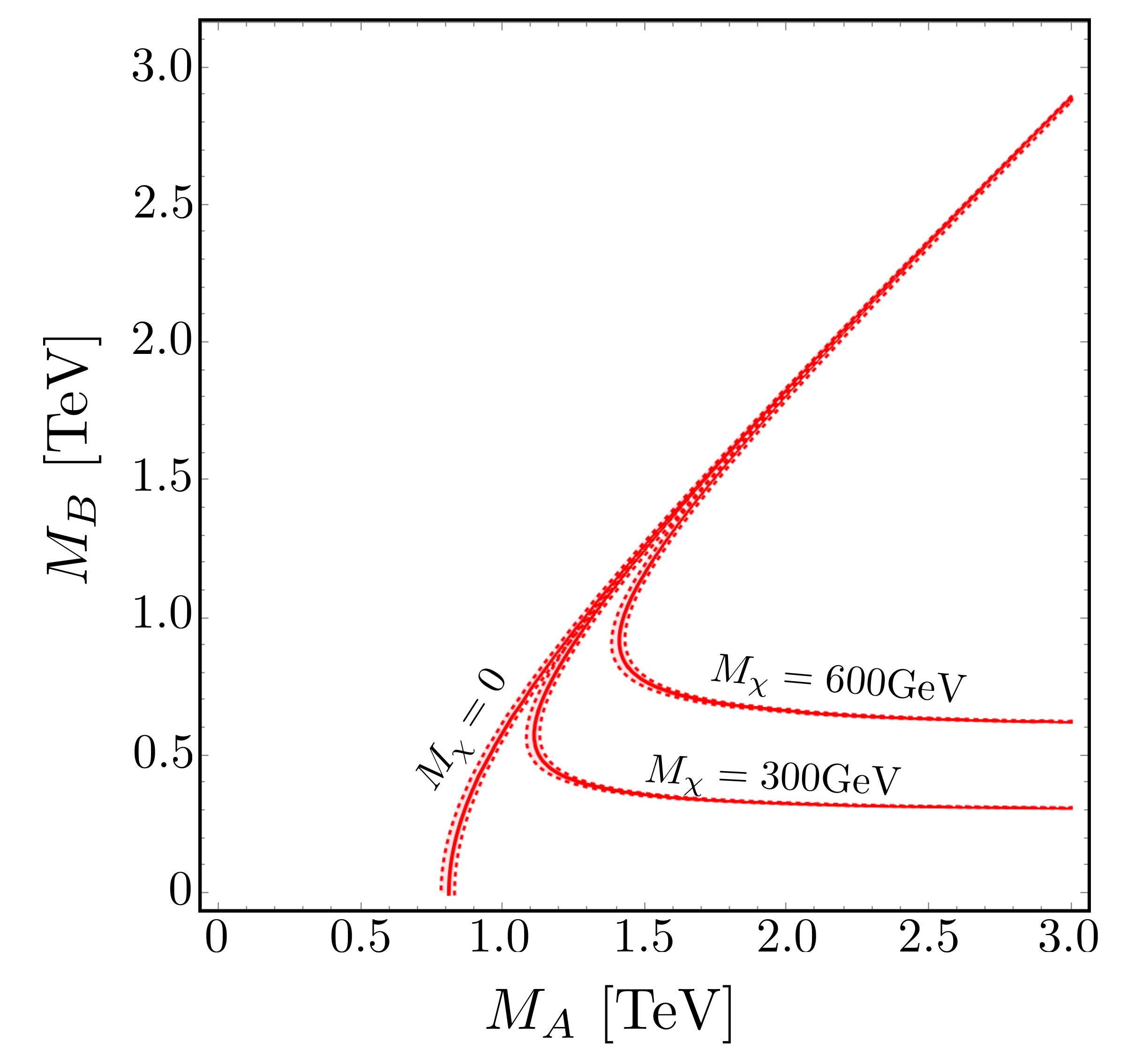} &
\includegraphics[clip=true, trim=0cm -0.1cm 0cm 0.7cm, height=4.2cm, width=4.3cm]{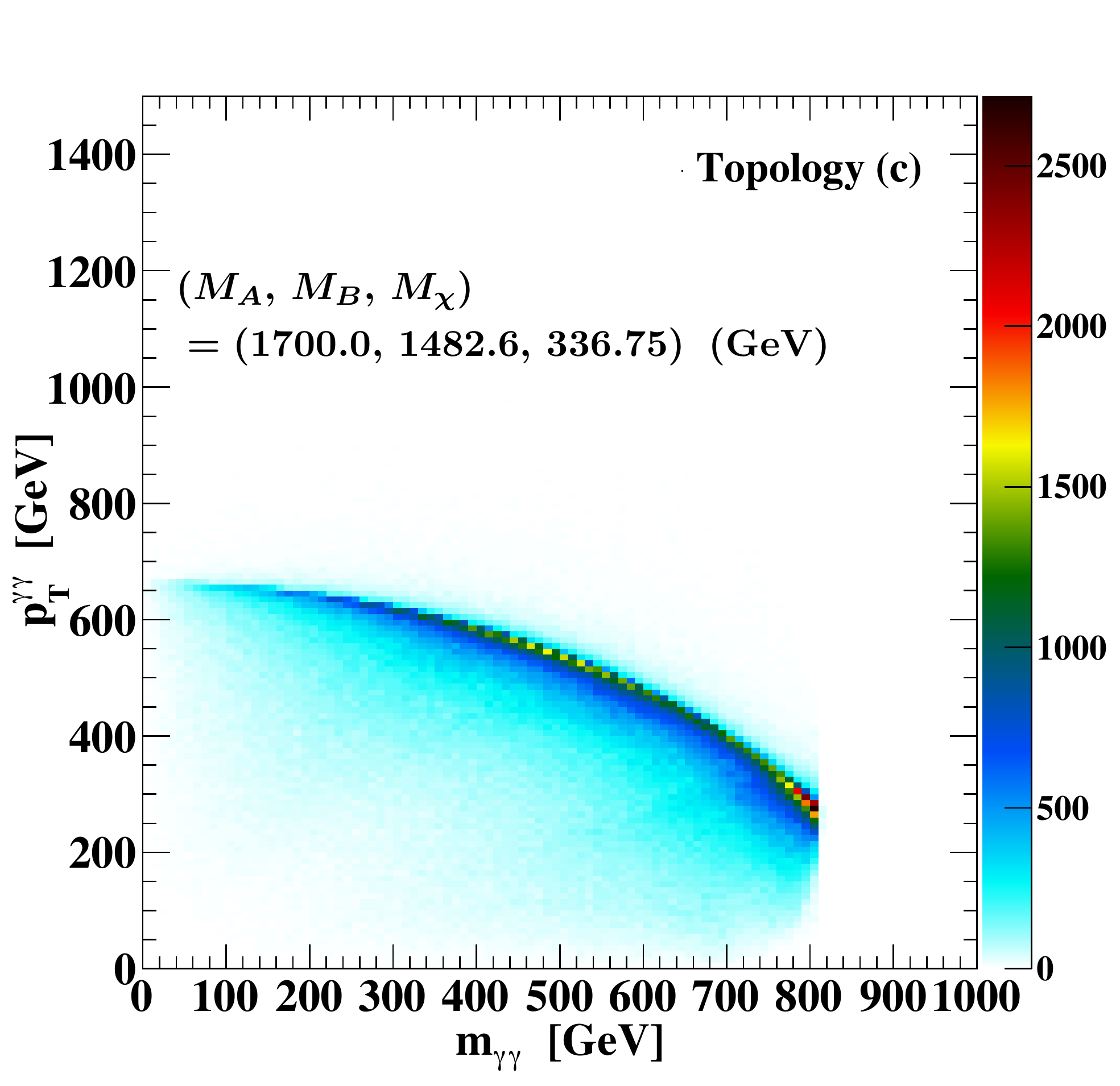}
\end{tabular}
\caption{\label{fig:massspectrum} Left panels: the allowed mass regions at $1\sigma$, selected by the best-fit.
Right panels: temperature plots showing the correlation between $m_{\gamma\gamma}$ and $p_T^{\gamma\gamma}$,
using parton-level Monte Carlo events with a representative mass spectrum consistent 
with the best-fit values in Eqs.~\eqref{etaE_fit} and \eqref{E}.
}
\end{figure}

\paragraph*{{\bf Discussions and outlook.}}
Since the number of experimentally measurable parameters for the antler topology is two
(namely,  $\eta$ and $E$)~\cite{Cho:2012er}, the underlying mass spectrum is not fully determined.
However, a phenomenologically motivated scenario is the case where the decay is symmetric,
i.e., $B_1=B_2$ and $\chi_1=\chi_2$. We then have three input mass parameters, two of which can be
given as functions of the third mass, using the measured values for $\eta$ and $E$.
Taking $M_{\chi}$ as a free parameter, we find that $M_A$ and $M_B$ can be expressed as follows:
\bea
M_B&=&\left(e^{-\eta/2}E+\sqrt{e^{-\eta}E^2+4M_{\chi}^2}\,\right)/2\,, \\
M_A&=&\sqrt{2M_B^2(\cosh\eta+1)}\,.
\eea
The upper-left panel of Fig.~\ref{fig:massspectrum} displays the corresponding
1$\sigma$ mass ranges for the $A$ (blue region and curves) and $B$ (red region and curves)
particles as a function of $M_{\chi}$.

For the sandwich topology of Fig.~\ref{fig:diagram}(b), we can similarly reduce the number of input mass
degrees of freedom by considering the simple case of $\chi_1=\chi_2$ as a well-motivated phenomenological scenario.
Then, using the measurements (\ref{etaE_fit}),
we can predict the masses of two of the unknown particles, say $M_{B_1}$ and $M_{B_2}$,
as a function of the other two, $M_{A}$ and $M_{\chi}$, as shown in
the middle left panel of Fig.~\ref{fig:massspectrum} ($M_{B_1}$ only for illustration).

Finally, for the two-step cascade topology of Fig.~\ref{fig:diagram}(c),
only one parameter, Eq.~(\ref{E}), can be measured from the data.
This provides one relation among the three unknown masses $M_A$, $M_B$ and $M_\chi$,
which is depicted in the bottom left panel of Fig.~\ref{fig:massspectrum}.

\begin{figure}[t]
\centering
\begin{tabular}{cc}
\includegraphics[trim=0.5cm 0 0.7cm 0, width=4.2cm]{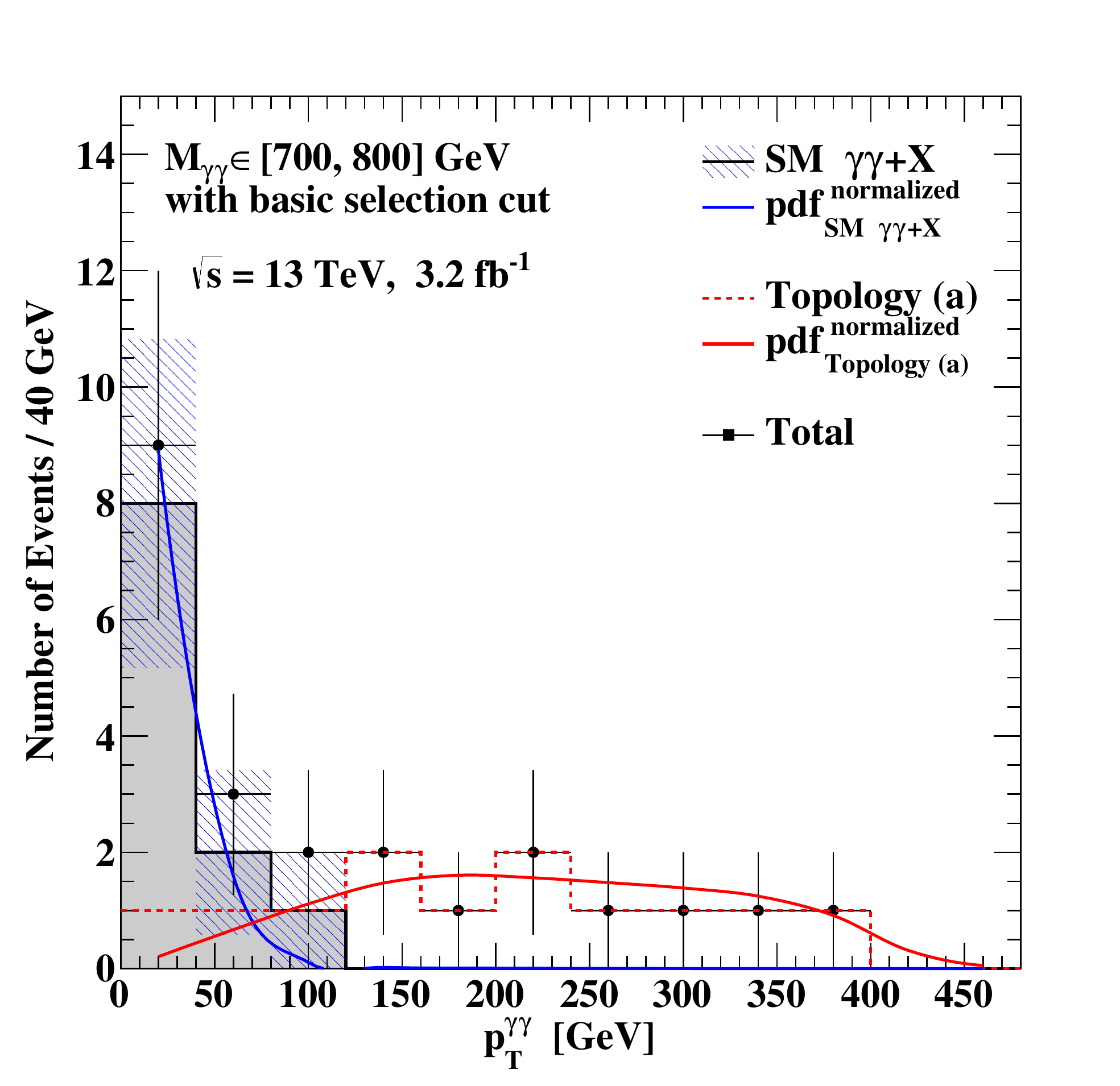} &
\includegraphics[trim=0.5cm 0 0.7cm 0, width=4.2cm]{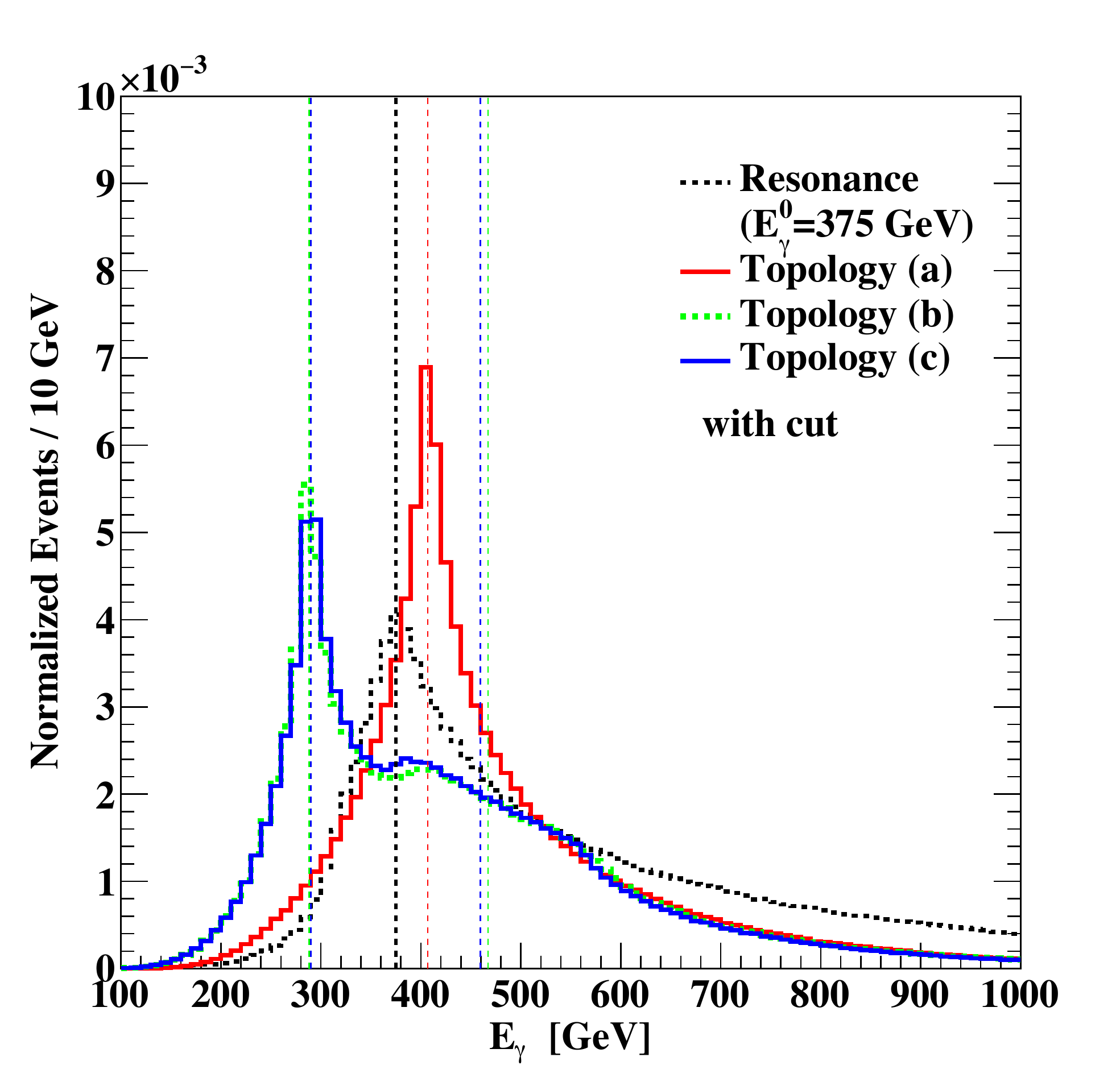}
\end{tabular}
\caption{\label{fig:discriminator}
Left: $p_T^{\gamma\gamma}$ distribution of events (in a single pseudo-experiment) 
near the bump, $m_{\gamma\gamma}\in (700,800)$ GeV,
for the event topology of Fig.~\ref{fig:diagram}(a), 
with the mass spectrum from Fig.~\ref{fig:massspectrum}.
Solid lines show the expected distributions.  
Right: Unit-normalized photon energy distributions for
the conventional scenario with a heavy resonance of mass $750$ GeV and width $45$ GeV (black dotted line),
and the three cascade decay scenarios: the antler topology (red solid), the sandwich topology (blue solid), and the two-step cascade (green dotted).
The dashed vertical lines mark the expected energy-peaks.
}
\end{figure}

We have seen that the cascade event topologies from Fig.~\ref{fig:diagram}
can provide a good fit to the diphoton invariant mass spectrum in Fig.~\ref{fig:fitting}.
It is therefore natural to ask, what other kinematic variables of the diphoton 
system can be used to test our hypothesis. One such possibility is the transverse 
momentum of the diphoton system, $p_T^{\gamma\gamma}$, since it is sensitive to
other objects recoiling against the two photons. However, there exists an inverse correlation
between the two diphoton kinematic variables, $m_{\gamma\gamma}$ and $p_T^{\gamma\gamma}$,
as illustrated in the right panels of Fig.~\ref{fig:massspectrum}:
events with extreme values of $m_{\gamma\gamma}$ have relatively small 
$p_T^{\gamma\gamma}$ and vice versa. The anti-correlation trend is especially pronounced for the
antler event topology, as demonstrated in the left panel of Fig.~\ref{fig:discriminator}, 
where we show the $p_T^{\gamma\gamma}$ distribution of simulated events near the bump,
$m_{\gamma\gamma}\in (700,800)$ GeV, for 3.2 fb$^{-1}$ of data with the ATLAS selection cuts
\cite{ATLAS}. For such events, the typical angular separation (in the laboratory frame)
between the two photons is anticipated to be large, and if the photons are almost back-to-back, 
then so must be the two $\chi$'s, yielding a relatively small net $p_T^{\gamma\gamma}$.
Fig.~\ref{fig:discriminator} is consistent with Ref.~\cite{Moriond} and shows that the signal events 
lead to a rather featureless tail in the $p_T^{\gamma\gamma}$ distribution. With the accumulation of more data, 
$p_T^{\gamma\gamma}$ will eventually be a good discriminator between the conventional resonance scenario 
(with relatively soft $p_T^{\gamma\gamma}$)
and the cascade decay scenarios considered here.

Another handle to discriminate among the competing interpretations of Fig.~\ref{fig:diagram} is
provided by the photon energy spectrum.
In the conventional case of a single resonance with a large decay width \cite{higgs,susy,ed,sd,eft},
the photon energy spectrum has a single peak at half the resonance mass~\cite{Agashe:2012bn,1971NASSP.249.....S},
which may show a sharp kink structure if the heavy resonance is singly produced~\cite{Chen:2014oha}.
On the other hand, the energy distribution for the (symmetric) antler scenario develops a peak at a {\it different} position,
\bea
E_{\gamma}=(M_B^2-M_{\chi}^2)/(2M_B)\,.
\eea
For the other two cases, the corresponding photon spectrum could develop a {\it double-bump} 
structure depending on the underlying mass spectrum~\cite{Agashe:2013eba}.
These expectations are summarized in the right panel of Fig.~\ref{fig:discriminator}.

As the excess was observed in the {\it inclusive} diphoton channel \cite{ATLAS,CMS:2015dxe,Moriond}, 
we have focused our attention primarily on the kinematics of the diphoton system itself.
Of course, more {\it exclusive} studies could target the detector signatures of the additional 
particles $\chi_i$. For example, if the particles $\chi_i$ are stable and weakly interacting,
they will be invisible in the detector and cause missing transverse energy $\met$. The predicted
$\met$ distribution would be similar to the $p_T^{\gamma\gamma}$ distribution shown in Fig.~\ref{fig:discriminator},
and at this point seems to be disfavored by the data \cite{Moriond} 
(another constraint would be provided by the inclusive diphoton plus $\met$ search for new physics \cite{Aad:2015hea}).
Of course, the particles $\chi_i$ could be visible, or further decay visibly themselves. 
The exact nature of their signatures (and kinematic distributions) is rather model-dependent 
and beyond the scope of this letter.

Finally, we note the potential impact of spin correlations on our analysis.
It is well-known that the overall shape of invariant mass distributions can be distorted by
the introduction of non-trivial spin correlations \cite{Wang:2006hk,Burns:2008cp}.
One could then repopulate most of the signal events in a (relatively) narrow region around the peak,
which would further improve the fit. Let $f_S(m)$
be the relevant $m_{\gamma\gamma}$ distribution in the presence of spin correlations.
For the antler and sandwich cases, one can write \cite{Agashe:2010gt,Edelhauser:2012xb}
\bea
f_S(m) \sim \left\{
\begin{array}{l l}
m(c_1+c_2 t+c_3 t^2) \, ,& 0 \leq m \leq e^{-\eta}E, \\ [1mm]
m[c_4+c_5 t+c_6 t^2 & \\
+(c_7+c_8 t+c_9 t^2)\ln t] \, ,& e^{-\eta}E \leq m \leq E.
\end{array}\right.  \label{eq:fs}
\eea
Here $t\equiv m^2/E^2$ and $c_i$ ($i=1, \ldots, 9$) represent coefficients encoding the underlying spin information.
For the decay topology in Fig.~\ref{fig:diagram}(c), the relevant expression is given by
the first line of Eq.~\eqref{eq:fs}~\cite{Agashe:2010gt,Wang:2006hk}:
\begin{equation}
f_S(m) \sim m(d_1+d_2 t+d_3 t^2)\, \hbox{ for }\, 0\leq m \leq E\,,
\end{equation}
and the presence of the additional terms beyond Eq.~(\ref{eq:cascadef})
can also favorably sculpt the distribution in the vicinity of the peak.

In conclusion, we investigated the nature of the anomalous excesses reported by the ATLAS and CMS Collaborations
in terms of cascade decay topologies from a heavy, possibly quite narrow, resonance. Our scenarios can generically
accommodate a (relatively) large width of the peak accompanied with a (relatively) small diphoton transverse momentum.
We also discussed the potential of distinguishing the competing interpretations with more data,
using the diphoton transverse momentum and photon energy distributions.
We eagerly await the resolution of this puzzle with new data from the LHC.


\section*{Acknowledgments}
We thank Peisi Huang and Carlos Wagner for insightful discussions.
DK, KK and JCP thank IBS-CTPU for hospitality and supports during the completion of this work. DK and KM thank the organizers of the Miami 2015 conference where significant portion of this work was completed.
This work is supported by NSF (PHY-0969510), DOE (DE-FG02-12ER41809, DE-SC0007863), IBS (IBS-R018-D1), and the Korean Ministry of Education (NRF- 2013R1A1A2061561).


\begin{thebibliography}{999}


\bibitem{ATLAS}
  ATLAS Collaboration [ATLAS Collaboration],
  ATLAS-CONF-2015-081.

\bibitem{CMS:2015dxe}
  CMS Collaboration [CMS Collaboration],
  CMS-PAS-EXO-15-004.

\bibitem{Moriond}
M.~Delmastro, ``Diphoton searches in ATLAS", and 
P.~Musella,  ``Diphoton Searches in CMS",
talks given at the 51st Rencontres de Moriond EW 2016,
La Thuile, March 17th, 2016,
\verb^https://indico.in2p3.fr/event/12279^


\bibitem{higgs}
  A.~Angelescu, A.~Djouadi and G.~Moreau,
  arXiv:1512.04921 [hep-ph];
  S.~D.~McDermott, P.~Meade and H.~Ramani,
  arXiv:1512.05326 [hep-ph];
  A.~Kobakhidze, F.~Wang, L.~Wu, J.~M.~Yang and M.~Zhang,
  arXiv:1512.05585 [hep-ph];
  R.~Martinez, F.~Ochoa and C.~F.~Sierra,
  arXiv:1512.05617 [hep-ph];
  D.~Becirevic, E.~Bertuzzo, O.~Sumensari and R.~Z.~Funchal,
  arXiv:1512.05623 [hep-ph];
  W.~Chao, R.~Huo and J.~H.~Yu,
  arXiv:1512.05738 [hep-ph];
  P.~Agrawal, J.~Fan, B.~Heidenreich, M.~Reece and M.~Strassler,
  arXiv:1512.05775 [hep-ph];
  Y.~Bai, J.~Berger and R.~Lu,
  arXiv:1512.05779 [hep-ph];
  S.~Ghosh, A.~Kundu and S.~Ray,
  arXiv:1512.05786 [hep-ph].
  R.~Benbrik, C.~H.~Chen and T.~Nomura,
  arXiv:1512.06028 [hep-ph];
  G.~Li, Y.~Mao, Y.~Tang, C.~Zhang, Y.~Zhou and S.~Zhu,
  arXiv:1512.08255 [hep-ph].

  
\bibitem{susy}
  C.~Petersson and R.~Torre,
  arXiv:1512.05333 [hep-ph];
  S.~V.~Demidov and D.~S.~Gorbunov,
  arXiv:1512.05723 [hep-ph];
  E.~Gabrielli, K.~Kannike, B.~Mele, M.~Raidal, C.~Spethmann and H.~Veerm\"{a}e,
  arXiv:1512.05961 [hep-ph];
  L.~M.~Carpenter, R.~Colburn and J.~Goodman,
  arXiv:1512.06107 [hep-ph].


\bibitem{ed}
  P.~Cox, A.~D.~Medina, T.~S.~Ray and A.~Spray,
  arXiv:1512.05618 [hep-ph];
  A.~Ahmed, B.~M.~Dillon, B.~Grzadkowski, J.~F.~Gunion and Y.~Jiang,
  arXiv:1512.05771 [hep-ph];
  E.~Megias, O.~Pujolas and M.~Quiros,
  arXiv:1512.06106 [hep-ph].


\bibitem{sd}
  K.~Harigaya and Y.~Nomura,
  arXiv:1512.04850 [hep-ph];
  Y.~Nakai, R.~Sato and K.~Tobioka,
  arXiv:1512.04924 [hep-ph];
  E.~Molinaro, F.~Sannino and N.~Vignaroli,
  arXiv:1512.05334 [hep-ph];
  S.~Matsuzaki and K.~Yamawaki,
  arXiv:1512.05564 [hep-ph];
  J.~M.~No, V.~Sanz and J.~Setford,
  arXiv:1512.05700 [hep-ph];
  L.~Bian, N.~Chen, D.~Liu and J.~Shu,
  arXiv:1512.05759 [hep-ph];
  J.~S.~Kim, J.~Reuter, K.~Rolbiecki and R.~R.~de Austri,
  arXiv:1512.06083 [hep-ph].

    
\bibitem{eft}
  Y.~Mambrini, G.~Arcadi and A.~Djouadi,
  arXiv:1512.04913 [hep-ph];
  M.~Backovic, A.~Mariotti and D.~Redigolo,
  arXiv:1512.04917 [hep-ph];
  D.~Buttazzo, A.~Greljo and D.~Marzocca,
  arXiv:1512.04929 [hep-ph];
  A.~Pilaftsis,
  arXiv:1512.04931 [hep-ph];
  R.~Franceschini {\it et al.},
  arXiv:1512.04933 [hep-ph];
  S.~Di Chiara, L.~Marzola and M.~Raidal,
  arXiv:1512.04939 [hep-ph];
  T.~Higaki, K.~S.~Jeong, N.~Kitajima and F.~Takahashi,
  arXiv:1512.05295 [hep-ph];
  J.~Ellis, S.~A.~R.~Ellis, J.~Quevillon, V.~Sanz and T.~You,
  arXiv:1512.05327 [hep-ph];
  M.~Low, A.~Tesi and L.~T.~Wang,
  arXiv:1512.05328 [hep-ph];
  B.~Bellazzini, R.~Franceschini, F.~Sala and J.~Serra,
  arXiv:1512.05330 [hep-ph];
  R.~S.~Gupta, S.~J\"{a}ger, Y.~Kats, G.~Perez and E.~Stamou,
  arXiv:1512.05332 [hep-ph];
  B.~Dutta, Y.~Gao, T.~Ghosh, I.~Gogoladze and T.~Li,
  arXiv:1512.05439 [hep-ph];
  Q.~H.~Cao, Y.~Liu, K.~P.~Xie, B.~Yan and D.~M.~Zhang,
  arXiv:1512.05542 [hep-ph];
  S.~Fichet, G.~von Gersdorff and C.~Royon,
  arXiv:1512.05751 [hep-ph];
  D.~Curtin and C.~B.~Verhaaren,
  arXiv:1512.05753 [hep-ph];
  J.~Chakrabortty, A.~Choudhury, P.~Ghosh, S.~Mondal and T.~Srivastava,
  arXiv:1512.05767 [hep-ph];
  C.~Csaki, J.~Hubisz and J.~Terning,
  arXiv:1512.05776 [hep-ph];
  A.~Falkowski, O.~Slone and T.~Volansky,
  arXiv:1512.05777 [hep-ph];
  D.~Aloni, K.~Blum, A.~Dery, A.~Efrati and Y.~Nir,
  arXiv:1512.05778 [hep-ph];
  A.~Alves, A.~G.~Dias and K.~Sinha,
  arXiv:1512.06091 [hep-ph].
  
  
  


\bibitem{Knapen:2015dap}
  S.~Knapen, T.~Melia, M.~Papucci and K.~Zurek,
  arXiv:1512.04928 [hep-ph];
  J.~Bernon and C.~Smith,
  arXiv:1512.06113 [hep-ph].


\bibitem{Datta:2005zs} 
  A.~Datta, K.~Kong and K.~T.~Matchev,
  Phys.\ Rev.\ D {\bf 72}, 096006 (2005)
  [Phys.\ Rev.\ D {\bf 72}, 119901 (2005)]
  doi:10.1103/PhysRevD.72.096006, 10.1103/PhysRevD.72.119901
  [hep-ph/0509246].

\bibitem{Macesanu:2002ew} 
  C.~Macesanu, C.~D.~McMullen and S.~Nandi,
  Phys.\ Lett.\ B {\bf 546}, 253 (2002)
  doi:10.1016/S0370-2693(02)02694-1
  [hep-ph/0207269].
  
\bibitem{DiazCruz:2003bx} 
  J.~L.~Diaz-Cruz, D.~K.~Ghosh and S.~Moretti,
  Phys.\ Rev.\ D {\bf 68}, 014019 (2003)
  doi:10.1103/PhysRevD.68.014019
  [hep-ph/0303251];
  J.~D.~Mason, D.~E.~Morrissey and D.~Poland,
  Phys.\ Rev.\ D {\bf 80}, 115015 (2009)
  doi:10.1103/PhysRevD.80.115015
  [arXiv:0909.3523 [hep-ph]].
  
      
\bibitem{Han:2009ss}
  T.~Han, I.~W.~Kim and J.~Song,
  Phys.\ Lett.\ B {\bf 693}, 575 (2010)
  doi:10.1016/j.physletb.2010.09.010
  [arXiv:0906.5009 [hep-ph]].

\bibitem{Agashe:2010gt}
  K.~Agashe, D.~Kim, M.~Toharia and D.~G.~E.~Walker,
  Phys.\ Rev.\ D {\bf 82}, 015007 (2010)
  doi:10.1103/PhysRevD.82.015007
  [arXiv:1003.0899 [hep-ph]].

\bibitem{Cho:2012er}
  W.~S.~Cho, D.~Kim, K.~T.~Matchev and M.~Park,
  Phys.\ Rev.\ Lett.\  {\bf 112}, no. 21, 211801 (2014)
  doi:10.1103/PhysRevLett.112.211801
  [arXiv:1206.1546 [hep-ph]].




\bibitem{Edelhauser:2012xb}
  L.~Edelhauser, K.~T.~Matchev and M.~Park,
  JHEP {\bf 1211}, 006 (2012)
  doi:10.1007/JHEP11(2012)006
  [arXiv:1205.2054 [hep-ph]].

\bibitem{Alwall:2014hca} 
  J.~Alwall {\it et al.},
  JHEP {\bf 1407}, 079 (2014)
  doi:10.1007/JHEP07(2014)079
  [arXiv:1405.0301 [hep-ph]].

\bibitem{Sjostrand:2006za} 
  T.~Sjostrand, S.~Mrenna and P.~Z.~Skands,
  JHEP {\bf 0605}, 026 (2006)
  doi:10.1088/1126-6708/2006/05/026
  [hep-ph/0603175].
  
\bibitem{deFavereau:2013fsa} 
  J.~de Favereau {\it et al.} [DELPHES 3 Collaboration],
  JHEP {\bf 1402}, 057 (2014)
  doi:10.1007/JHEP02(2014)057
  [arXiv:1307.6346 [hep-ex]].

  
  

\bibitem{Agashe:2012bn}
  K.~Agashe, R.~Franceschini and D.~Kim,
  Phys.\ Rev.\ D {\bf 88}, no. 5, 057701 (2013)
  [arXiv:1209.0772 [hep-ph]].

\bibitem{1971NASSP.249.....S}
  F.~W. {Stecker}, ``{Cosmic gamma rays},'' {\it NASA Special Publication}
  {\bfseries 249} (1971).

\bibitem{Chen:2014oha}
  C.~Y.~Chen, H.~Davoudiasl and D.~Kim,
  Phys.\ Rev.\ D {\bf 89}, no. 9, 096007 (2014)
  [arXiv:1403.3399 [hep-ph]].

\bibitem{Agashe:2013eba}
  K.~Agashe, R.~Franceschini and D.~Kim,
  JHEP {\bf 1411}, 059 (2014)
  [arXiv:1309.4776 [hep-ph]].


\bibitem{Aad:2015hea} 
  G.~Aad {\it et al.} [ATLAS Collaboration],
  Phys.\ Rev.\ D {\bf 92}, no. 7, 072001 (2015)
  doi:10.1103/PhysRevD.92.072001
  [arXiv:1507.05493 [hep-ex]].
  
  

\bibitem{Wang:2006hk}
  L.~T.~Wang and I.~Yavin,
  JHEP {\bf 0704}, 032 (2007)
  doi:10.1088/1126-6708/2007/04/032
  [hep-ph/0605296].

\bibitem{Burns:2008cp}
  M.~Burns, K.~Kong, K.~T.~Matchev and M.~Park,
  JHEP {\bf 0810}, 081 (2008)
  doi:10.1088/1126-6708/2008/10/081
  [arXiv:0808.2472 [hep-ph]].

\end{thebibliography}
\end{document}